\documentclass[12pt,a4paper]{article}
\usepackage[english]{babel}
\usepackage{graphicx}

\newcommand{\alt}{\mathbin{\lower 3pt\hbox
   {$\rlap{\raise 5pt\hbox{$\char'074$}}\mathchar"7218$}}}
\newcommand{\agt}{\mathbin{\lower 3pt\hbox
   {$\rlap{\raise 5pt\hbox{$\char'076$}}\mathchar"7218$}}}

\begin{document}
\setcounter{footnote}{0}
\setcounter{equation}{0}
\setcounter{figure}{0}
\setcounter{table}{0}
%\vspace*{5mm}

\begin{center}
{\large\bf
Possibility of the 2D Anderson Transition \\
and Generalized Lyapunov Exponents}

\vspace{4mm}
I. M. Suslov \\
P.L.Kapitza Institute for Physical Problems,
\\ 119337 Moscow, Russia \\
\vspace{6mm}

\begin{minipage}{135mm}
{\rm {\quad}  The possible existence of the Anderson transition in
2D systems without interaction and spin-orbit effects (such as the
usual Anderson model) becomes recently a subject of controversy in
the literature \cite{1,0,3}. Comparative analysis of approaches
based on generalized Lyapunov exponents is given, in order to
resolve controversy. }
\end{minipage} \end{center}

\vspace{6mm}
\begin{center}
{\bf 1. Introduction} \\
\end{center}

Possible existence of the Anderson transition in the 2D case
becomes recently a subject of controversy in the literature.
Kuzovkov~et~al~\cite{1} studying  a growth of the second moments
for a particular solution of the quasi-1D Schroedinger equation
and interpreting results in terms of the signal theory came to
conclusion that the first order Anderson transition exists in the
usual 2D Anderson model. This incredible result was analyzed by
Markos~et~al~\cite{0}  and the opposite conclusion was drawn:
behavior of the second moments is qualitatively different from
found in \cite{1} and contains no evidence of the metallic phase.
On the other hand, the present author \cite{3} used the analogous
approach and came to conclusion on existence of the
Kosterlitz-Thouless type transition, which however can be realized
not in all 2D systems.

The present paper has an aim to analyze the existing controversy
and is organized in the following manner. In Sec.\,2 we give brief
exposition of paper \cite{3}, which we consider as the proper
treatment of the problem. Approach by Kuzovkov~et~al~\cite{1}
is discussed in Sec.\,3: the results for the second moments are
shown to be partially correct but their interpretation
is not satisfactory. Section 4 deals with the approach advanced
by Markos~et~al~\cite{0}: their statements are shown to be
related with improper calculation of the matrix product.
Conclusions are made in Sec.\,5.

%\newpage
\vspace{4mm}
\begin{center}
{\bf 2. Second moments of the Cauchy solution \\
and their relation  with the Anderson transition} \\
\end{center}

Consider the 2D Anderson model described by the discrete Schroedinger
equation
$$
\psi_{n+1,m}+\psi_{n-1,m}+\psi_{n,m+1}+\psi_{n,m-1}+ V_{n,m} \psi_{n,m} =
E\psi_{n,m} \eqno(1)
$$
and interpret it as a recurrence relation in the variable $n$,
which we accept as a longitudinal coordinate.  Initial conditions
are assumed to be fixed on the left end of the system, while the
periodic boundary conditions are accepted in the transverse
direction, $\psi_{n,m+L}=\psi_{n,m}$.  Site energies $V_{n,m}$ are
considered as uncorrelated random quantities with the first two
moments
$$
\langle\,V_{n,m} \,\rangle=0\,,\qquad
\langle\,V_{n,m} V_{n',m'} \,\rangle = W^2 \delta_{n,n'} \delta_{m,m'} \,.
\eqno(2)
$$
The growth of the second moments for this problem can be studied using the old
idea by Thouless \cite{5} based on the observation that
variables $\psi_{n,m}$ are statistically independent of $V_{n,m}$
with the same $n$. The main quantity of interest is
$\langle \psi_{n,m}^2\rangle$;  solving (1) for
 $\psi_{n+1,m}$ and averaging its square, we can relate
it with the pair correlators containing lower values of $n$.
Deriving analogous equations for the pair
correlators, we end with the closed system of difference equations for the
quantities
$$
x_{m,m'}(n)\equiv \langle\,\psi_{n,m}\psi_{n,m'} \,\rangle\,,
$$
$$
\,\,y_{m,m'}(n)\equiv \langle\,\psi_{n,m}\psi_{n-1,m'} \,\rangle\,,
\eqno(3)
$$
$$
z_{m,m'}(n)\equiv \langle\,\psi_{n-1,m}\psi_{n,m'} \,\rangle \,,
$$
which for $E=0$  has a form \cite{3}
$$
x_{m,m'}(n+1)=W^2 \delta_{m,m'} x_{m,m'}(n) +x_{m+1,m'+1}(n)+
x_{m-1,m'+1}(n)+ x_{m+1,m'-1}(n) +x_{m-1,m'-1}(n) +
$$
$$
 +x_{m,m'}(n-1)
+y_{m+1,m'}(n) +y_{m-1,m'}(n) +z_{m,m'+1}(n)+ z_{m,m'-1}(n)
\eqno(4)
$$
$$
y_{m,m'}(n+1)=- x_{m+1,m'}(n) -x_{m-1,m'}(n)- z_{m,m'}(n)
\phantom{nnnnnnnnnnnnnnnnnnnnnnnnnnnnnnnnnnnnnnnnnnnnnnnnnnmmmmmmmmmmmmmm}
$$
$$
z_{m,m'}(n+1)=- x_{m,m'+1}(n) -x_{m,m'-1}(n)- y_{m,m'}(n)\,.
\phantom{nnnnnnnnnnnnnnnnnnnnnnnnnnnnnnnnnnnnnnnnnnnnnnnnnnmmmmmmmmmmmmmm}
$$
This is a set of the linear difference equations with independent
of $n$ coefficients and its solution is exponential in $n$
\cite{48}
$$
x_{m,m'}(n) =x_{m,m'} {\rm e}^{\,\beta n}\,,\qquad
y_{m,m'}(n) =y_{m,m'} {\rm e}^{\,\beta n}\,,\qquad
z_{m,m'}(n) =z_{m,m'} {\rm e}^{\,\beta n}\,.\qquad
\eqno(5)
$$
The formal change of variable is useful
$$
x_{m,m'} \equiv \tilde x_{m,m'-m}= \tilde x_{m,l} \,,\qquad
{\rm etc.,} \eqno(6)
$$
where $l=m'-m$. Substitution of (6) and (5) to
(4) gives the difference equations whose coefficients contain no
 $m$ dependence and their solution is exponential in $m$
 $$
\tilde x_{m,l} = x_{l} \,{\rm e}^{\,i\,p m}\,,\qquad  {\rm etc.,}
\eqno(7)
$$
where allowed values for $p$  are
determined by the periodical boundary conditions in the
transverse direction.
Excluding  $y_{m,l}$ and $z_{m,l}$ from the first equation in (40), we end
with the equation
$$
x_{l+2}\,{\rm e}^{-i\,p}+x_{l-2}\,{\rm e}^{i\,p}+ V \delta_{l,0} x_{l} =
\epsilon x_{l} \,,\qquad x_{l+L}=x_l \,,
\eqno(8)
$$
$$
\epsilon= 2 \cosh{\beta}\,,\qquad
V=\frac{W^2 \sinh{\beta}}{\cosh\beta-\cos{p}}
$$
describing a single impurity in a periodic chain.
The positive exponents $\beta_s$ for finite odd $L$ are determined by
equation\,\footnote{\,There are also pure imaginary $\beta_s$,
which are inessential for us. The detailed form of solution for
$x_{m,m'}(n)$ is given by Eq.44 in \cite{3} and demonstrates
equality of localization lengths in transversal and longitudional
directions.}
$$
2(\cosh{\beta_s}-\cos{p_s} )=W^2 \coth(\beta_s L/2) \,,
\qquad p_s=2 \pi s/L\,, \quad s=0,1,\ldots,L-1 \,.
\eqno(9)
$$
Allowed values of $p_s$ and $\beta_s$ become dense in the
large $L$ limit, and the quantities $\beta$ and
$p$ can be considered as continuous;
the minimal value of $\beta$ is realized for $p=\pi$ and can be
easily found in the large $L$ limit
$$
\beta_{min}= \left\{
\begin{array}{cc} \cosh^{-1} \left(W^2/2-1 \right)\,, & W^2>4\\ {
}\\ \displaystyle\frac{2}{L} \tanh^{-1} \left(W^2/4 \right)\,, &
W^2<4  \\ {   }\\ \displaystyle\frac{2\ln{L} -2\ln\ln{L}+\ldots
}{L} \,, & W^2=4 \,.  \end{array} \right.
\eqno(10)
$$
The character of solution is qualitatively changed at $W_c=2$;
this value of disorder corresponds to a singular point found
in \cite{1} but its physical sense is not
evident.\,\footnote{\,Analogous calculations for $E\ne 0$
give $W_c^2=2\sqrt{4-E^2}$ in accordance with \cite{1}. }

Interpretation of results can be given using the finite-size
scaling approach in the form given by Pichard and Sarma
\cite{7}\,\footnote{\,In comparison with \cite{7a}, it is based
on more general motivation and is not restricted by one-parameter
scaling hypothesis.}. A quasi-1D system is always localized and
the localization length $\xi_{1D}$ can be introduced for it. The
knowledge of this length can be used for the study of the
Anderson transition in the higher-dimensional systems. It is
easy to show that  $\xi_{1D}\to const$ in the localized phase and
$\xi_{1D}/L\to \infty$ in the metallic state in the large $L$
limit \cite{3,7}. If we introduce a scaling parameter
$$
g(L)=\frac{\xi_{1D}}{L}
\eqno(11)
$$
then it increases with $L$ in the metallic phase and decreases
in the insulator phase. In the framework of one-parameter scaling
hypothesis a following relation can be postulated \cite{7a}
$$
g(L)= F\left(\frac{L}{\xi} \right).
\eqno(12)
$$
In this case, $g(L)$ remains constant in the critical point
and its general behavior is shown in Fig.\,1,a. The relation (12)
was never proved but its validity can be expected by analogy with
the usual phase transition theory.

\begin{figure}
\centerline{\includegraphics[width=6.1 in]{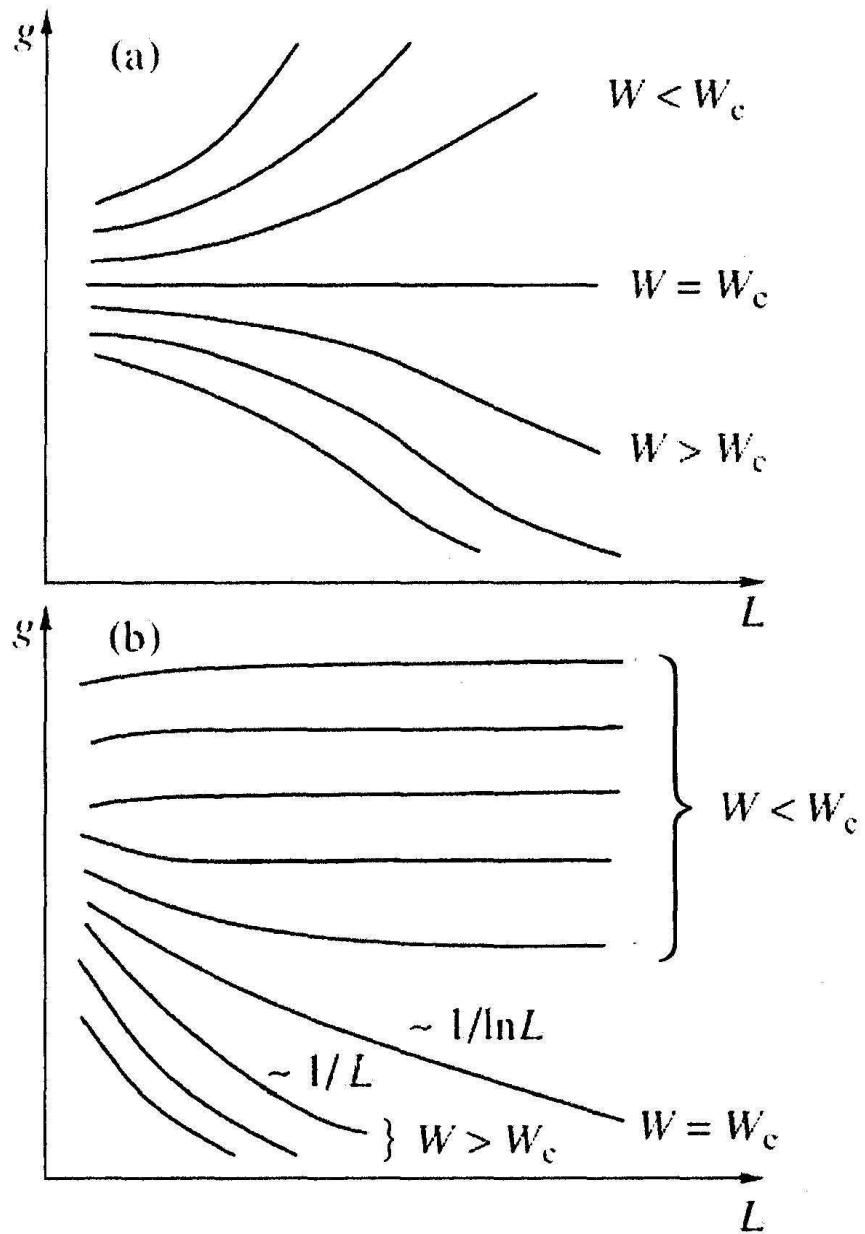}} \caption{ a
---  Typical behavior of $g(L)$ in the case of one-parameter
scaling: different curves for $W>W_c$ or $W<W_c$ can be made
coinciding by a scale transformation. b --- Behavior of $g(L)$
according to Eq.\,9: crude violation of scaling is evident.
 } \label{fig1}
\end{figure}

The solution $\psi_{n} (r_{\bot})$
of the Cauchy problem for the quasi-1D  Schroedinger
equation with the initial conditions on the left edge
allows decomposition
$$
\psi_{n} (r_{\bot}) = A_1 h_{n}^{(1)} (r_{\bot})\, {\rm e}^{\,\gamma_1 n}
                +A_2 h_{n}^{(2)} (r_{\bot})\, {\rm e}^{\,\gamma_2 n}  +\ldots
                +A_m h_{n}^{(m)} (r_{\bot}) \,{\rm e}^{\,\gamma_m
        n} \, \eqno(13)
        $$
where $r_{\bot}$ is the
transverse coordinate (like $m$ in (1)), the quantities
$h_{n}^{(s)} (r_{\bot})$ have no systematic growth in $n$, while
the Lyapunov exponents $\gamma_s$ tend to the constant values in
the large $n$ limit. According to Mott \cite{8},
eigenfunctions of a quasi-1D system can be constructed by
matching two solutions of the type (13)  increasing from two
opposite edges of the system.  The tails of the
eigenfunction will be determined by the minimal positive Lyapunov
exponent $\gamma_{min}$ and $\xi_{1D}$ can be estimated as
$1/\gamma_{min}$; this estimation in combination with the scaling
relation (12) is a basis for the most popular of contemporary
numerical algorithms (see \cite{7a} and a review article
\cite{9}). Due to evident analogy between (5) and (13) we can
refer (as suggested in \cite{1}) to the exponents $\beta_s$ as
generalized Lyapunov exponents. It is easy to show \cite{3}, that
positive $\beta_s$ and positive $\gamma_s$ are in one to one
correspondence: a term containing $\beta_s$ is an averaged square
of the term containing $\gamma_s$. The usual correspondence
between the mean square and the most probable value gives
inequality $\beta_s\ge 2\gamma_s$ \cite{3}.

Though no rigorous relation exists between $\beta_{min}$ and
$\gamma_{min}$,  these quantities are very close from
the physical viewpoint. Indeed, according to \cite{3}:
(a) inequality $\beta_{min}\ge 2 \gamma_{min}$ can be rigorously
proven; (b) the order of magnitude relation $\beta_{min}\sim
\gamma_{min}$ takes place in the typical physical
situation\,\footnote{This relation follows from existence
of the log-normal distribution and the fact that fluctuations of
$\gamma_s$ are of the same order (or less) as their mean value.
These properties can be proved in the limits of weak and
strong disorder and are confirmed by extensive numerical studies
in the intermediate region (see references in \cite{3,4}).};
(c) $\beta_{min}$ and $\gamma_{min}$ are practically
equivalent from viewpoint of one-parameter scaling philosophy,
and relations
$$
\frac{1}{\gamma_{min}L} = F\left(\frac{L}{\xi} \right)
\qquad {\rm and} \qquad
\frac{1}{\beta_{min}L} = F\left(\frac{L}{\xi} \right)
\eqno(14)
$$
can be postulated on the same level of rigorousness.

If the correlation length $\xi_{1D}$ is estimated as $1/\beta_{min}$,
then the behavior of $g(L)$ determined by Eq.\,9 has a form
presented in Fig.\,1,b; one can see an essential
difference from a typical scaling situation (Fig.\,1,a).
Since there is no growth of $g(L)$ for
large $L$,  the state with long-range order (i.e. metallic phase)
is absent, in accordance with \cite{15}.  Exponential
localization takes place for $W>W_c$, while the behavior specific
for a critical point is realized in all range $W<W_c$.  The
latter situation corresponds to localization with the divergent
correlation length and probably should be interpreted as  power
law localization.  The transition at $W=W_c$ is of the
Kosterlitz-Thouless type and should not be mixed with the usual
Anderson transition. This result can explain the observable 2D
metal-insulator transition \cite{11}\,\footnote{\,According
to Last and Thouless \cite{101}, the hopping conductivity for
power-localized states goes to zero for $T\to 0$ as a power of
$T$. Consequently, decrease of
resistivity with growth of $T$ is not so quick as for
exponentially localized states and can be changed to increase
by  other effects (for example, by $T^2$ contribution from
electron-electron scattering). Such picture is close to
observable in experiment \cite{102}.}
and does not imply a serious revision in the
weak localization region.

Violation of scaling relation (12) is clear in Fig.\,1,b
since $g(L)$ is not constant for $W=W_c$. It is
still more evident for $W<W_c$, when different curves have different
constant limits for $L\to\infty$ and surely cannot be
%matched
made coinciding by a scale transformation.

One may suspect that these results are related with
our estimation of $\xi_{1D}$ as $1/\beta_{min}$, since
the quantities $\beta_{min}$ and $\gamma_{min}$ can be
essentially different.
In fact, inequality $\beta_{min}\ge 2\gamma_{min}$ is sufficient
for existence of phase transition. Indeed, one can see from
this inequality that $\gamma_{min} \to 0$ for $ L\to\infty$,
if $ W<W_c$; on the other hand, for large $W$ existence of
exponential localization is beyond any doubt and finiteness
of $\gamma_{min}$ is evident; it can be rigorously proved in the
large $W$ limit \cite{100}.  Of course, the upper bound
for $\gamma_{min}$ does not forbid it to decrease more rapidly than $1/L$,
as it should be for a true metallic state. But such possibility is
reliably excluded by numerical studies.

The latter fact, in combination with inequality
$\beta_{min}\ge 2\gamma_{min}$, is sufficient to claim violation
of scaling for $\gamma_{min}$. Indeed, for $W=W_0<W_c$ the scaling
parameter $1/\gamma_{min}L$ cannot increase and is bounded from
below by constant $2/\beta_{min}L$. If we take $W_1<W_0$ such
that $\beta_{min}(W_1)<2\gamma_{min}(W_0)$, then
$\gamma_{min}(W_1)<\gamma_{min}(W_0)$ and the constant limit
of the scaling parameter $1/\gamma_{min}L$ cannot be the
same  for different $W$: we return to the picture presented in
Fig.1,b.  As a result, substitution of $\gamma_{min}$ for
$\beta_{min}$ does not lead to qualitative changes in the
presented picture and inequality $\beta_{min}\ge 2\gamma_{min}$
is sufficient for the most responsible statements.
%, i.e. existence
%of phase transition and violation of one-parameter scaling
%\cite{3}.

One can see that interpretation in terms of finite size
scaling leads to unambiguous conclusion on  existence
of the 2D phase transition. This statement
does not contradict to numerical results, if the raw data
are considered \cite{3}. The opposite conclusion made by numerical
researchers is based on interpretation in terms of one-parameter
scaling, which is inadmissible here. Recent numerical
results clearly demonstrate (see Fig.\,37 in \cite{9}) that
possibility of the 2D phase transition cannot be rejected on
numerical grounds.

However,  we do not consider the conventional variant of finite
size scaling as indisputable. Calculation of
Lyapunov exponents is not equivalent to diagonalization of the
Hamiltonian: for example, statistical independence of
$\psi_{n,m}$ and $V_{n,m'}$  is valid in the first but not
the second case, and one can suspect oversimplification
of the problem. For reliable estimation of $\xi_{1D}$ one
needs detailed study of the coefficients $A_i$ in Eq.13
appearing in Mott's construction for eigenfunctions.
In the general case, relation
$\xi_{1D}\sim 1/\gamma_{min}$ can be violated and some effective
exponent $\gamma_{eff}$ should be used instead $\gamma_{min}$
\cite{3}. If the scaling relation of type (14) is postulated for
$\gamma_{eff}$, then its dependence on parameters is determined by
scaling itself. Analysis shows possibility of two variants: (a)
the 2D phase transition is removed; (b) the 2D transition remains,
though behavior of the correlation length becomes different.  One
can suggest that both possibilities are realized in different
models.

The described approach can be generalized to higher
dimensionality \cite{2,4}. Unfortunately, the analytical results
for the critical disorder appear to be in essential contradiction
with corresponding numerical results. The interpretations of this
fact can be different, but in any case it is related with crude
violation of one-parameter scaling \cite{4}.

%\newpage

%\vspace{6mm}
\begin{center}
{\bf 3. Approach by Kuzovkov~et~al~\cite{1}} \\
\end{center}

Let us discuss the difference of approach presented in Sec.\,2 from
one suggested by Kuzovkov~et~al~\cite{1,2}.  The initial system
of equations (4) and its higher dimensional analogue (Eq.\,5 in
\cite{4}) coincide with those used in \cite{1,2}. However, the
quantity $z_{mm'}(n)$ was not introduced in \cite{1,2} and its
role was played by $y_{m'm}(n)$. As a result, the system of equations
had no complete difference form and could not be solved in the natural
manner with evaluation of full spectrum of exponents $\beta_s$.
Instead, the authors of \cite{1,2} used the $Z$-transform and
introduced the so called filter function $H(z)$. The latter, as
was demonstrated on simple examples, has the poles
corresponding to eigenvalues of the transfer matrix,
and, in principle,
this approach allows to find the full
spectrum of generalized Lyapunov  exponents. However,
some problems arise in the practical realization of this scheme:

(a) The solution was obtained only in the thermodynamic limit
$L\to\infty$, and the problems exist with interpretation of
results (see below).

(b) Averaging over translations in the transversal direction was
used for simplification of the problem. This procedure is
disputable since it can lead to elimination of some
singularities of $H(z)$. Indeed, translational invariance for the
solution (7)  of Eq.\,4  and its higher dimensional analog
(Eq.\,10 in \cite{4}) is absent and averaging over translations
eliminates all terms with transverse momentum $\bf p\ne 0$.
Analogously, averaging of the squared solution eliminates all
terms with $\bf p\ne 0,\, G/2$ where $\bf G$ is a vector of a
reciprocal lattice corresponding to the main diagonal of the
Brillouin zone. We can suggest by comparing results (though cannot
prove it rigorously) that the filter function $H(z)$ is analogous
to $\psi_{nm}^2$ and only terms with $\bf p=0$ and $\bf p=G/2$ are
essential for it. Fortunately for the authors of \cite{1,2}, a
condition $\bf p=G/2$ corresponds to the minimal exponent
$\beta_{min}$  for $d=2$ and $d\ge 4$, so the critical values
$\sigma'_0$ (corresponding to $W_c$ in \cite{3,4}) were found
correctly for these cases. In the case $d=3$, the exponent
$\beta_{min}$  does not correspond to $\bf p=G/2$ and  a zero
value for the critical disorder was obtained  in \cite{2} for the
band center $E=0$, in a striking contrast with \cite{4}.

Attempt to justify this point made in the paper \cite{12}
is not convincing, in our opinion. In fact, the results reduce
to the statement that averaging procedure does not eliminate
the maximum exponent $\beta_{max}$, corresponding to $\bf p=0$.
This statement is surely correct but it has no relation
to physics of the problem.

(c) In the general case, function $H(z)$ can have not only poles
corresponding to eigenvalues of the transfer matrix but also
another singularities, which are physically irrelevant.  In our
opinion, it is an origin of the second critical point $\sigma_0$
obtained in \cite{1,2}  for higher dimensionality; we see no
evidence for it in the spectrum of $\beta_s$.  Correspondingly,
 we see no evidence of a special role of dimensionality $d=6$,
which is surely absent in the exact field theory approach
\cite{6}.

\vspace{4mm}

The papers \cite{1,2} were formulated in the engineer's
language (using the concepts of signals, filters etc.),
which has no direct relation to the Anderson transition.
The authors relate the extended states with a stable filter
and the localized states with unstable filter, i.e.
introduce their own localization criterion whose correspondence
with conventional one was never studied.

The limit $L\to\infty$ was taken in \cite{1,2} from the
very beginning, and the finite-size scaling approach could
not be used for interpretation of results. The authors
interpreted the Anderson transition as being of the first
order, considering two branches of the filter function $H(z)$
as two different phases existing simultaneously; such
interpretation does not have clear physical sense.

%The limit $L\to\infty$ was taken in \cite{1,2} from the
%very beginning, and the finite-size scaling approach could
%not be used for interpretation of results.  As a consequence,
%the Anderson transition was interpreted as being of
%the first order. In our opinion, the authors had no serious
%grounds for such a statement\,\footnote{\,Two branches of the
%filter function $H(z)$ were interpreted as two different phases
%existing simultaneously; such interpretation looks rather vague.}
%and it was made simply as the
%only way to make contradiction with all available information
%not so striking.

%Nevertheless, p
Possibility of the first order transition is in
conflict with the old ideas by Mott \cite{8}: according to the
Poincare theorem, a small change in the energy or the disorder
strength induces small changes of the wave functions, and hence
the state of the system changes continuously\,\footnote{\,It does
not exclude that certain quantities can display a jump-like
behavior; the conductivity, according to Mott, belongs to such
quantities.}.  It is essential, that the Poincare theorem is valid
only for a finite system and allows existence of the mobility
edge, if the localization radius has divergency in it. Existence
of the divergent length is not characteristic for the first order
transition.

The possibility of power-law localization was also mentioned in
\cite{1,2}, but in the context, which is completely different
from that in  \cite{3,4}.

\vspace{6mm}
\begin{center}
{\bf 4. Approach by Markos~et~al~\cite{0}} \\
\end{center}

Markos~et~al~\cite{0} do not follow the natural procedure of
Sec.\,2, but advance more complicated approach. They
rewrite Eq.\,1 using the transfer matrix and construct the direct
product of two such matrices. After averaging, they arrive to a
linear system of equations determined by a matrix
$$
T=
  \left ( \begin{array}{cccc}
W^2\,1\bigotimes 1+ D_0 \bigotimes D_0 & -D_0\bigotimes 1 &
-1 \bigotimes D_0& 1 \bigotimes 1\\
D_0 \bigotimes 1 &  0 & -1 \bigotimes 1& 0 \\
1 \bigotimes D_0 & -1 \bigotimes 1&  0& 0\\
1 \bigotimes 1& 0& 0& 0
\end{array} \right) \,,
\eqno(15)
$$
where $D_0=E-H_0$ and $H_0$ is the Hamiltonian of the $n$-th slice for
a pure system, $1$ is the unit matrix of the size $L\times L$.
In principle, this approach is equivalent to that of Sec.\,2
but appears to be  practically untractable due to sophisticated
matrix constructions. The eigenvalues $\lambda=\exp(iq)$ of the
matrix (15) are declared to be determined by equation
$$
2 \cos{2q} - 2\kappa_i \kappa_j \cos{q} + (\kappa_i^2+ \kappa_j^2-2)
=2W^2 i\sin{q}
\eqno(16)
$$
where $\kappa_i=E-2\cos{p_i}$ and $p_i$ are allowed values of the
transverse momentum $p$. Equation (16) surely not coincides with
the corresponding Eq.\,9  in Sec.\,2. The main difference is the
absence of functions with the argument $qL$, which inevitably
arise due to the boundary conditions and can be absent only in
trivial cases. One can suspect, that this difference is related
with improper treatment of the disorder term (like $W^2 1\times 1$
in (15)), without which the problem is trivial. This term is local
in $m-m'$ (see (4)) and not diagonal in the momentum
representation, which is seemingly used in (16). It looks likely
that the local nature of this term was neglected and it was
replaced by a suitable constant. We can try such thing for the
system (4), replacing $\delta_{m,m'}$ by unity. Then (4) is solved
trivially
$$
x_{m,m'}(n) =x {\rm e}^{\,i p m+i p' m'}\, {\rm e}^{\,\beta n}\,,\qquad
y_{m,m'}(n) =y {\rm e}^{\,i p m+i p' m'}\, {\rm e}^{\,\beta n}\,,\qquad
z_{m,m'}(n) =z {\rm e}^{\,i p m+i p' m'}\, {\rm e}^{\,\beta n}\,,\qquad
\eqno(17)
$$
where allowed values for $p$ and $p'$  ($2 \pi s/L$, $s=0,1,\ldots,L-1$)
are determined by the boundary conditions and the quantities $x$, $y$, $z$
satisfy the equation
$$
  \left ( \begin{array}{ccc}
W^2 +4\cos{p}\cos{p'}-2\sinh{\beta} & 2\cos{p} & 2\cos{p'}\\
2\cos{p} & {\rm e}^\beta & 1  \\
2\cos{p'} &  1 &  {\rm e}^\beta
\end{array} \right) \,
  \left ( \begin{array}{c}
x\\y\\z
\end{array} \right) \, =0  \,.
\eqno(18)
$$
The determinant vanishes under condition
$$
2 \cosh{2\beta} - 2\kappa \kappa' \cosh{\beta} + (\kappa^2+ \kappa'^2-2)
=2W^2 \sinh{\beta}\,
\eqno(19)
$$
where $\kappa=-2\cos{p}$, $\kappa'=-2\cos{p'}$. In the case $E=0$
(when Eq.\,4 holds) this equation is identical to (16), if correspondence
$\beta=iq$ is taken into account. We see that, indeed, the disorder term
in (15) was treated inadequately and its local nature was neglected.
 Physically,  equation (16) corresponds not to the true
 Anderson model but to its degenerate version when site energies
 $V_{n,m}$ are independent of $m$.

In fact, the error is present already in the matrix (15). To obtain
the term with disorder,  one needs to produce averaging of the kind
$$
\left ( \begin{array}{cc}
V_1 & 0 \\
0 & V_2 \end{array} \right) \,
\bigotimes
\left ( \begin{array}{cc}
V_1 & 0 \\
0 & V_2 \end{array} \right) \,=
\left ( \begin{array}{cccc}
V_1 V_1 & 0 & 0 &0 \\
0 &  V_1 V_2 & 0 & 0 \\
0 & 0 &  V_2 V_1 & 0\\
0& 0& 0& V_2 V_2
\end{array} \right) \,
\longrightarrow W^2
\left ( \begin{array}{cccc}
1 & 0 & 0 &0 \\
0 & 0 & 0 & 0 \\
0 & 0 &  0 & 0\\
0& 0& 0& 1
\end{array} \right) \,
\eqno(20)
$$
and the result cannot be represented as $W^2\,1\bigotimes 1$. The
latter form is valid for $V_1=V_2$ in Eq.\,20, while in the general
case it corresponds to the model (1) with $V_{n,m}$ being independent
of $m$, in accordance with the previous analysis. Such model is
of no interest and all conclusions made in \cite{0} are irrelevant
for the problem under consideration.

\vspace{6mm}
\begin{center}
{\bf 5. Conclusion} \\
\end{center}

The following conclusions can be made from our analysis
of generalized Lyapunov exponents:

(a) A spectrum of $\beta_s$ allows complete analytical
investigation.

(b) Interpretation of results in terms of conventional variant of
finite-size scaling leads to unambiguous conclusion on existence
of the 2D transition of the Kosterlitz-Thouless type and crude
violation of one-parameter scaling in any dimensionality.

(c) Attempt to restore one-parameter scaling \cite{15} by
replacement $\gamma_{min}$ by $\gamma_{eff}$ eliminates the $2D$
transition from roughly  half of models.

(d) Results for the critical disorder obtained by
Kuzovkov~et~al~\cite{1,2}   are correct for $d=2$ and $d\ge 4$
(not for $d=3$), but their interpretation is not
satisfactory.

(e) Qualitatively different conclusions  by Markos et al
\cite{0} are based on improper calculation of the matrix product.

(f) Above conclusions do not contradict numerical results, if
the raw data are considered; interpretation of the latter in
terms of one-parameter scaling is inadmissible.

\vspace{3mm}

This work is partially supported by RFBR  (grant 06-02-17541).

%\newpage

\vspace{3mm}

\end{document}